\documentstyle[psfig,aps,preprint]{revtex}
\def\lsim{\mathrel{\rlap {\raise.5ex\hbox{$ < $}}
{\lower.5ex\hbox{$\sim$}}}}
\def\gsim{\mathrel{\rlap {\raise.5ex\hbox{$ > $}}
{\lower.5ex\hbox{$\sim$}}}}
\def\np#1#2#3{{\it {Nucl. Phys.}} {\bf{B#1}} (#2) #3}
\def\pl#1#2#3{{\it {Phys. Lett.}} {\bf{B#1}} (#2) #3}
\def\prl#1#2#3{{\it {Phys. Rev. Lett. }}{\bf{#1}} (#2) #3}
\def\pr#1#2#3{{\it {Phys. Rev.}} {\bf{D#1}} (#2) #3}

\begin{document}
\begin{titlepage}
\begin{flushright}

hep-ph/9710563{\hskip.5cm}\\
\end{flushright}
\begin{centering}
\vspace{.3in}
{\bf RADIATIVE GUT SYMMETRY BREAKING IN A $R$-SYMMETRIC FLIPPED $SU(5)$ MODEL
}\\
\vspace{2 cm}
{A. DEDES$^{1}$
 , C. PANAGIOTAKOPOULOS$^2$ and K. TAMVAKIS$^1$}\\
\vskip 1cm
{\it $^1$Division of Theoretical Physics,}\\
{\it University of Ioannina, GR-45110, Greece}\\ \vspace{0.5cm}
{\it $^2$Physics Division, School of Technology, University of
Thessaloniki, Greece}\\ \vspace{0.5cm}

%{$^1 $\it{Physics Department, University of Ioannina\\
%Ioannina, GR45110, GREECE}}\\
%
%\vskip 1cm
%{$^2$\it {School of Technology, University of Thessaloniki\\
%Thessaloniki, GREECE}}\\

\vspace{1.5cm}
{\bf Abstract}\\
We study the generation of the GUT scale through radiative corrections
in the context of a $R$-symmetric ``flipped" $SU(5)\times U(1)_X$ model.
A negative mass squared term for the GUT Higgs fields 
develops due to radiative effects along a flat direction
 at a superheavy energy scale. The $R$-symmetry is essential in
 maintaining triplet-doublet splitting and $F$-flatness in the
 presence of non-renormalizable terms. The model displays radiative electroweak
 symmetry breaking and satisfies all relevant phenomenological constraints.
\end{centering}
\vspace{.1in}

\vspace{1cm}
\begin{flushleft}

October 1997\\
\end{flushleft}
\hrule width 6.7cm \vskip.1mm{\small \small}
E-mails\,:\,adedes@cc.uoi.gr, costapan@eng.auth.gr, tamvakis@cc.uoi.gr

\end{titlepage}

{\bf  INTRODUCTION}

  The Standard Model (SM) and its $N=1$ supersymmetric extension\cite{NHK},
  the Minimal Supersymmetric
 Standard Model (MSSM), can be naturally embedded in a Grand Unified Theory
 \cite{GUTS} (GUT)
 with interesting phenomenological
 and cosmological consequences. GUTs can successfully predict the electroweak
 mixing angle $\sin^2\theta_W$, fermion
 mass relations as well as provide a mechanism for the creation of the baryon
  asymmetry of the Universe\cite{BA}. However, in the framework of quantum
field theory
  no severe restrictions exist on the gauge group or the field content of a GUT
  apart from the requirement that it should incorporate the Standard Model. Many
  possibilities are allowed including minimal $SU(5)$ and its
extensions\cite{SU5},
  varieties of $SO(10)$ \cite{SO10} and $E(6)$ \cite{E6} models e.t.c. A GUT can
be
in principle
  accommodated in the
  framework of superstring theory\cite{GSW}. The assumption
 that the GUT is the low
energy
  field theory limit of a four dimensional heterotic superstring
compactification
  imposes serious restrictions on the spectrum. For gauge groups realized at
  level $k=1$ of the World Sheet Affine Algebra, only the chiral multiplets in
  the vector and antisymmetric tensor
 representations of $SU(n)$ groups and the vector
  and spinor representations of $SO(n)$ groups are massless. The absence of
  adjoint scalars severely restricts the possibilities of breaking to the MSSM
  through the Higgs mechanism and diminishes the number of candidate GUT
models.
  Apart from these restrictions superstrings offer a new possibility. The GUT
  gauge group does not have to be simple in order to guarantee unification
  {\footnote{The term GUT now refers to a gauge group that only partially
   contains the
  SM gauge group}} .
  Semi-simple or product groups are equally acceptable since string theory
  takes over the task of gauge coupling unification. Among the few GUT examples
  embeddable in superstrings is the so-called ``flipped" $SU(5)\times U(1)$
  \cite{BN}
  model which has been explicitly constructed and studied in the framework
   of four
  dimensional free fermionic superstrings\cite{FFF}\cite{NA}\cite{JRKT}.
  When a GUT is considered in the
  context of string theory the GUT scale\cite{USSM} is distinct from the string
scale.
  Despite the fact that this is not necessarily a problem for product gauge
  groups it has been termed as the {\it {mismatch problem}}.

  Can
  the GUT scale be generated by radiative
  corrections in an analogous fashion to the generation of the electroweak
  breaking scale through dimensional transmutation in the MSSM ?
  The point of view realized in the present article is
     that the symmetry breaking scales associated with the effective
    field theories of a GUT or the Standard Model  are generated through
dimensional transmutation \cite{Col} \cite{Tamvakis}
     while the Planck scale $ M_P $  and the supersymmetry (SUSY)
     breaking scale $m_s$ are
     ``fundamental" and,
     presumably, accounted for by strings or non-perturbative physics.
    Although
    the idea of generating  the GUT scale through radiative
    corrections is not new, its existing realizations $\cite{Gato}$
are not satisfactory  for various reasons.
    These are, non-embeddability
    in strings, ``baroque" field content or lack of symmetries that could
    guarantee flatness in the presence of non-renormalizable corrections.
 In a recent paper H. Goldberg
$\cite{Goldberg}$ considers a gauge singlet superfield $S$ coupled to a
pair of adjoint fields in a supersymmetric $SU(5)$ model. In this model,
the soft breaking mass term of the gauge singlet $S$ becomes negative
and develops a vacuum expectation value (v.e.v)
$<S>$ which ultimately defines the GUT scale.

In the
  present article we are going to study 
   the generation of the GUT scale through radiative corrections
in a prototype $k=1$ string embeddable GUT.
These
  corrections, controlled by the supersymmetry breaking scale $m_s$ , can
  give rise to a logarithmically distant from $m_s$
scale $M_X$ lying close to the scale at which
  the soft SUSY-breaking squared masses of the GUT Higgs scalars become
  negative. Such a mechanism requires, of course,
 the existence of a D- and  F-flat direction for the relevant
  fields in the supersymmetric limit. As a prototype 
   GUT we shall employ a version of the ``flipped"
  $SU(5)\times U(1)_X$ model that satisfies the
    $k=1$ string embeddability criteria, possesses a discrete $R$-symmetry that
    guarantees triplet-doublet splitting and flatness along the direction
    responsible for the generation of 
     $M_X$, displays radiative electroweak symmetry breaking
    and satisfies all phenomenological low energy constraints.

{\bf 1.  A $R$-SYMMETRIC VERSION OF $SU(5)\times U(1)_X$ .}

   In the present article
  we shall work with a simple version of the flipped $SU(5)$ model possessing
  an almost
  minimal field content which, however, is sufficient
to  illustrate the mechanism under
  investigation for generating  the GUT scale.
  The minimal chiral superfield content of the flipped $SU(5)\times U(1)_X$
model
  consists of the {\it{matter superfields}}, in three family replicas
  \begin{equation}
F_i({\bf{10}},1/2)\,\,\,,\,\,\,\,f^c_i({\bf{\overline{5}}},-3/2)\,\,\,,\,\,\,\,
  l^c_i({\bf{1}},5/2)\,\,\,, 
\end{equation}
%%%%%%%%%%%%%%%%
the {\it{GUT Higgses}}

\begin{equation}
H({\bf{10}},1/2)\,\,\,,\,\,\,\,\overline{H}({\bf{\overline{10}}},-1/2)\,\,\,,
\end{equation}
  and the {\it{electroweak Higgses}}

\begin{equation}h({\bf{5}},-1)\,\,\,,\,\,\,\,\overline{h}({\bf{\overline{5}}},1)\,\,\,.
  \end{equation}

 We shall also introduce four additional gauge singlet superfields, three $N_i$'s and one
$\phi$ , and
 an additional pair of tenplets
 \begin{equation}
H'({\bf{10}},1/2)\,\,\,\,,\,\,\,\,\overline{H}'({\bf{\overline{10}}},-1/2)\,\,\,.
 \end{equation}
 All these fields are massless at the Planck scale. The masslessness of most of them
 will be
 protected by additional symmetries that will be shortly imposed. Nevertheless,
experience from the string
 model\cite{NA}\cite{JRKT} itself has shown that tree-level $O(M_P)$ mass terms
allowed by
symmetries are not always present.
 Adopting this point of view we shall assume  that $\phi$ and the additional
pair
  of tenplets $H'$ , $\overline{H}'$ have an intermediate scale mass despite
the fact that
  $O(M_P)$ masses for them are allowed by the symmetries.

  The $SU(5)\times U(1)_X$ gauge symmetry breaking at a superheavy scale
requires the existence of a
  F-flat direction for $H$ , $\overline{H}$ . In order to achieve F-flatness
one could impose
  discrete symmetries. However, with conventional discrete symmetries one can
hardly protect
  F-flatness from non-renormalizable terms. For example, the frequently imposed
discrete symmetry
  $H\rightarrow -H$ does not forbid
the dangerous
  superpotential term $(H\overline{H})^2$ which lifts the F-flatness and
forbids a v.e.v. $\sim 10^{16} GeV$ .
 Moreover, such a symmetry may generate a serious domain wall
problem in the early
  Universe. This problem becomes more severe if the phase transition associated
with the superheavy
  scale takes place at  temperatures $\sim m_s$ as is expected to be the
case in the
  context of superstring embeddable models. In contrast, $R$-symmetries are
capable
  of forbidding dangerous non-renormalizable terms to all orders. For example,
if we impose
  a $R$-symmetry under which $H$ and $\overline{H}$ transform trivially, all
terms
  of the type $(H\overline{H})^{n}$ are forbidden. Such a discrete symmetry is
not
  broken by a large v.e.v. of $H$ , $\overline{H}$ and the domain wall problem
has a
  good chance to be avoided. If the discrete symmetry is eventually
spontaneously broken, the
  domain wall problem could still be avoided provided the discrete $R$-symmetry
carries colour
  anomalies i.e. does not leave invariant the effective instanton vertex for QCD.
In this
  case the degeneracy between vacua separated by domain walls is lifted by QCD
instanton effects
  at temperatures of the order of the QCD scale (100 MeV). The resulting
pressure on the walls
  causes their collapse soon after the QCD phase transition
$\cite{Okun}$. This mechanism
assumes that the domain
  wall system does not dominate the energy density of the Universe before its
collapse.
  If the anomalous discrete $R$-symmetry breaks at the electroweak scale, this
condition
  is readily satisfied $\cite{Lazarides}$.
 In what follows we shall construct a version of the
flipped $SU(5)$ model
possessing such an anomalous discrete $R$-symmetry which is broken at the
electroweak scale.

   Consider a ${\cal{Z}}_3$ $R$-symmetry under which the fields transform as
 \begin{equation}\{ F,\,\,\,\, f^c,\,\,\,\,l^c,\,\,\,\,H,\,\,\,\,\overline{H}\}
\rightarrow
 \{ F,\,\,\,\,f^c,\,\,\,\,l^c,\,\,\,\,H,\,\,\,\,\overline{H} \} \end{equation}
 \begin{equation}\{ h,\,\,\,\,\overline{h},\,\,\,\,N \} \rightarrow
e^{\frac{2\pi i}{3}}\{
 h,\,\,\,\,\overline{h},\,\,\,\,N \}\end{equation}
 \begin{equation}\{\phi,\,\,\,\,H',\,\,\,\,\overline{H}'\}\rightarrow
  e^{-\frac{2\pi i}{3}}\{\phi,\,\,\,\,H',\,\,\,\,\overline{H}'\}\end{equation}
 and the {\it{superpotential}} as
 \begin{equation}{\cal{W}}\rightarrow e^{ \frac{2\pi
i}{3}}{\cal{W}}\;.
\end{equation}
 The above transformations  actually refer to the bosonic
components
 of the corresponding superfields. The fermionic components transform
 with an additional factor of $e^{\frac{2\pi i}{3}}$ relative to the bosonic
 ones. Consequently the
effective
 instanton QCD vertex is multiplied by a factor $e^{\frac{2\pi i}{3}}$ under a
discrete
 $R$-symmetry transformation which means that the
 discrete $R$-symmetry carries QCD  anomalies.

  In addition to the ${\cal{Z}}_3$ $R$-symmetry we also impose a ${\cal{Z}}_2$
{\it {matter
  parity}} under which the only fields transforming non-trivially are
   the matter fields and the three
singlets $N$

  \begin{equation}\{F,\,\,\,\,f^c,\,\,\,\,l^c,\,\,\,\,N,\,\,\,H',
\,\,\,\,\overline{H}'\}\rightarrow
-\{F,\,\,\,\,f^c,
  \,\,\,\,l^c,\,\,\,\,N,\,\,\,H',
\,\,\,\,\overline{H}'\}\;.
\end{equation}
  The matter parity singles out $\phi$ among the four singlets as the only one
   allowed to
  acquire an electroweak scale v.e.v., and generate the $\mu$-term. It also
  forbids $\phi$ to participate in the see-saw mechanism.
   Moreover, the
${\cal{Z}}_2$
  matter parity generates a cold dark matter candidate, the
  lightest supersymmetric particle, which is necessary
given that
  neutrinos are superlight ($m_{\nu}\sim M_W^3/M_X^2$).

  The renormalizable part of the superpotential respecting the
symmetry $SU(5)\times U(1)_X\times {\cal{Z}}_2 \times {\cal{Z}}_3$ is
\begin{equation}{\cal{W}}\sim FFh+Ff^c\overline{h}+f^cl^ch+HHh+
\overline{H}\overline{H}\overline{h}+F\overline{H}N+\phi h \overline{h}
+N^2\phi+ F\overline{H}' \phi 
+H'{\overline{H}}'+\phi^2\;.
\end{equation} Assuming that the extra pair of tenplets $H'$
,
$\overline{H}'$ as well as the singlet $\phi$ remain massless at
 the Planck scale, they obtain intermediate scale masses $\sim M_X^2/M_P$
 through the
non-renormalizable terms
$(H\overline{H})H'\overline{H}'$ , $H{\overline{H}}\phi^2$ .
 Note that, in general, 
supergravity
corrections generate a SUSY-breaking, $R$-symmetry-breaking tadpole
$\cite{Bagger}$
\begin{equation}m_s^2M_P(\phi+\phi^{*})\;.
\end{equation} As a result, a v.e.v for
$\phi$ is
induced 
\begin{equation}\langle \phi \rangle \sim M_P (m_s/m_{\phi})^2 \;,
\end{equation}
 where $m_{\phi}$ is the mass of $\phi$.  Demanding that $\langle
\phi \rangle
\sim m_s$ , we obtain  \begin{equation}m_{\phi}^2
\sim m_s M_P \;.
\end{equation}
This takes care of the $\mu$- problem.
It should be noted that due to the imposed symmetries the presence of
non-renormalizable terms does not affect neither the triplet-doublet splitting
 nor the F-flatness.
 Also the model possesses
  a mechanism to generate a $\mu$-term,
   provided the $R$-symmetry breaking is $\sim m_s$ .
    These are virtues in themselves which are worth emphasizing
    independently of the GUT scale generation realized by the model.

{\bf 2. RADIATIVE CORRECTIONS AND THE RGEs.}

  As already emphasized the superpotential $\cal{W}$ allows a large v.e.v.
  along the $F$-flat and $D$-flat direction $|H|=|{\overline{H}}|$ for the
  SM-singlets $N_H^c$ , ${\overline{N}}_{\overline{H}}^c$
 in $H$ , $\overline{H}$,
  respectively whose value $V_X$ is not determined at the tree-level. Such a
  v.e.v. breaks the $SU(5)\times U(1)_X$ gauge symmetry down to the SM gauge
  group $SU(3)_c\times SU(2)_L\times U(1)_Y$ . The colour triplets
  $D_H^c({\bf{3}}, {\bf{1}}, 1/3)$ ,
 ${\overline{D}}_{\overline{H}}^c({\bf{\overline{3}}},
  {\bf{1}}, -1/3)$ that survive the Higgs phenomenon pair up with the
  colour triplets $d_h({\bf{\overline{3}}}, {\bf{1}}, -1/3)$ ,
  ${\overline{d}}_{\overline{h}}({\bf{3}},
 {\bf{1}}, 1/3)$ in the Higgs pentaplets to
  form states with masses $\sim M_X \sim V_X$ through the superpotential
  couplings $HHh$ and ${\overline{H}}{\overline{H}}{\overline{h}}$ . The
  model exhibits triplet-doublet splitting which, however, is to a
  large extent a consequence of the imposed discrete symmetries.

    The GUT scale $M_X$ is defined as the scale at which the breaking
$SU(5)\times
    U(1)_X \rightarrow SU(3)_c\times SU(2)_L\times U(1)_Y$ occurs\footnote{
    The relation imposed by the $E_6$-normalization of $U(1)_X$ in flipped
$SU(5)$ is
    $$25\alpha_1^{-1}(M_X)=\alpha_5^{-1}(M_X)+24\alpha_X^{-1}(M_X)$$}
      or, equivalently,
    the scale at which the gauge couplings $\alpha_3$ and $\alpha_2$ meet, i.e.
    \begin{equation}\alpha_3(M_X)=\alpha_2(M_X)= \alpha_5(M_X)\equiv
\alpha_G\;.
\end{equation}
    One of the aims of the present article is to show how radiative corrections
    determine the value $M_X$ of this scale.
     Above $M_X$ the $SU(5)$ and $U(1)_X$ gauge couplings $\alpha_5$ and
    $\alpha_X$ converge and eventually
     meet at some scale $M_s$ which will be taken to be
the
    {\it{string unification scale}} $M_s\sim 5\times 10^{17} GeV$
    \begin{equation}\alpha_5(M_s)=\alpha_X(M_s)\equiv \alpha_{SU}\;.
\end{equation}
    The Renormalization Group Equations (RGEs) for the gauge couplings,
     apart from the standard
     minimal set of fields, involve the pair of extra tenplets
      $H'$ , ${\overline{H}}'$ as well. For their masses
       $\sim M_X^2/M_P$  we shall adopt a
phenomenological
     attitude and adjust them in the range
$10^{11}-10^{12} GeV$ in
     order to obtain an optimal fit of the low energy data.

      Let us now consider the RGEs for the
       gauge couplings. The leading logarithmic radiative corrections to the
       various parameters of the model are represented by the scale dependence
of
       the running parameters that satisfy the RGEs.
        We shall assume that the only appreciable couplings in
       the superpotential are the following :
       \begin{eqnarray}{\cal{W}}_1&=&\frac{1}{8}\lambda\epsilon^{AB\Gamma\Delta
E}
 H_{AB}H_{\Gamma\Delta}h_E +
\frac{1}{8}\overline{\lambda}\epsilon_{AB\Gamma\Delta E}
\overline{H}^{AB}\overline{H}^{\Gamma\Delta}\overline{h}^E
+Y_t^{ij}F^i_{AB}f^{cA}_j\overline{h}^B\nonumber \\[3mm]&+&
\frac{1}{2}Y_N^iF^i_{AB}\overline{H}^{AB}N_i \,\,\,\,(i=1,..3)\;.
\label{eq:superpotential}
\end{eqnarray}
The relevant soft-SUSY breaking terms corresponding to ${\cal{W}}_1$ are
\begin{eqnarray}
-{\cal L}_{soft}&=&m_{\cal H}^2 |{\cal H}|^2+
m_{\cal\overline{H}}^2 |{\cal\overline{H}}|^2+
m_h^2 |h|^2+m_{\overline{h}}^2 |\overline{h}|^2+
m_{{\cal F}_i}^2|{\cal F}_i|^2+m_{{f^c_i}}^2|f^c_i|^2
\nonumber \\[3mm]&+&
m_{{\cal N}_i}^2|{\cal N}_i|^2+
(\frac{1}{8}A_{\lambda}\lambda{\cal H}^2 h+
\frac{1}{8}\overline{A_{\lambda}}\overline{\lambda}\overline{\cal H}^2
\overline{h}+A_t^{ij}Y_t^{ij}{{\cal F}_i}f^c_j\overline{h}+
\frac{1}{2}A_N^iY_N^i{{\cal F}_i}\overline{{\cal H}}{{\cal N}_i}
\nonumber \\[3mm]&+&
\frac{1}{2}M_5\lambda_5\lambda_5+\frac{1}{2}M_1\lambda_1\lambda_1+\;h.c.)
\end{eqnarray}
%%%%%%%%%%%%%%%%%
where ${\cal H}$, ${\cal\overline{H}}$, $h$, $\overline{h},
{\cal F},
f^c, {\cal N}$ are the
scalar components of the superfields $H$, $\overline{H}$, $h$, $\overline{h},
F, f^c, N$, respectively and $\lambda_5$, $\lambda_1$ the gauginos of 
$SU(5)$ and $U(1)_X$, respectively.

As we shall explain shortly,
we are interested in the evolution of the soft SUSY-breaking masses from
$M_s$ to $M_X$ and more specifically in the RGEs
{\footnote {We
consider the third generation of the Yukawa couplings $Y_t^{ij}$
($Y_t^{33}\equiv Y_t$).
 The quantity $\cal{S}$ plays no role in the evolution and can be ignored. }}
  for $m_{\cal{H}}^2$ and $m_{\overline{\cal{H}}}^2$ :
%%%%%%%%%%%%%%%%%%%%%%%%
%%%%%%%%%%%%%%%%%%%%%
\begin{eqnarray}
Q\frac{dm_{{\cal H}}^2}{dQ}&=&\frac{1}{8 \pi^2}\{
3 \lambda^2 [m_h^2+2 m_{{\cal H}}^2+A_{\lambda}^2]\nonumber \\[3mm]&-&
\frac{72}{5}M_5^2g_5^2-\frac{1}{10}M_1^2 g_X^2+
\frac{1}{4} g_X^2 {\cal S}\} \label{soft1} \\[3mm]
%%%%%%%%%%%%%%%%%%%%%%%
Q\frac{dm_{{\cal\overline{H}}}^2}{dQ}&=&\frac{1}{8 \pi^2}\{
3 \overline{\lambda}^2 [m_{\overline{h}}^2+2 m_{{\cal \overline{H}}}^2+
\overline{A_{\lambda}}^2]+\sum_{i=1}^3 {Y^i}_N^2
 [m_{{\cal F}_i}^2+m_{{\cal \overline{H}}}^2+
m_{{\cal N}_i}^2+A_{N_i}^2]\nonumber \\[3mm]&-&
\frac{72}{5}M_5^2g_5^2-\frac{1}{10}M_1^2 g_X^2-
\frac{1}{4} g_X^2 {\cal S}\}\;,
\label{soft2}
\end{eqnarray}
%%%%%%%%%%%%%%%%%%%
where
%%%%%%%%%%%
\begin{equation}
{\cal S}=(m_{{\cal H}}^2-m_{{\cal\overline{H}}}^2)
-(m_h^2-m_{\overline{h}}^2)
-\frac{3}{2}\sum_{i=1}^3
m_{f^c_i}^2+\sum_{i=1}^3 m_{{\cal F}_i}^2\;.
\end{equation}
%%%%%%%%%%%%%%
Consider the flat direction that allows for a non-zero v.e.v. in the
supersymmetric limit.
The soft SUSY-breaking mass terms induce a small deviation from 
flatness which, with the
leading logarithmic corrections present in the 1-loop effective potential
for $\cal{H}$ ,
 $\overline{\cal{H}}$ taken into account, are,
in
 principle, able to generate a minimum at a scale $V_X\sim M_X$
with depth $\sim m_s^2 V_X^2$.
 . If this is the case, the GUT scale can be thought of
as 
  generated through radiative corrections. In order to investigate this
phenomenon it would be
  sufficient to consider the tree-level potential, given essentially by the
soft mass terms only, and
  study the renormalization group evolution of its parameters. Let us start at
high energies with
  positive soft masses-squared for the relevant fields
  $\cal{H}$ , $\overline{\cal{H}}$ , and
  assume that at some lower energy $Q_0$ a reversal
of sign occurs for
them{\footnote{
Consider a toy model with gauge group $U(1)$ involving the
superfields $\phi(1)$, $\overline{\phi}(-1)$, 
$f(-\frac{1}{2})$, $\overline{f}(\frac{1}{2})$.
We impose a ${\cal Z}_2$ symmetry under which
$f\rightarrow -f$ and a ${\cal Z}_4$ R-symmetry
under which $(f,\overline{f})\rightarrow i\;(f,\overline{f})$,
with the superpotential ${\cal W}\rightarrow -{\cal W}$.
Then, the renormalizable part of ${\cal W}$ is
%%%%%%%%%%%%%%%
\begin{eqnarray}
{\cal W}=\frac{1}{2}\lambda\;ff\phi +\frac{1}{2}\overline{\lambda}\; 
\overline{f}\overline{f}\overline{\phi}\;.
\nonumber
\end{eqnarray}
%%%%%%%%%%%%%%
The potential of the model in the supersymmetric limit
is given by
%%%%%%%%%%%%%%%
\begin{eqnarray}
V_1&=&|\lambda|^2 |f|^2\;(\frac{1}{4}|f|^2 + |\phi|^2)+
|\overline{\lambda}|^2 |\overline{f}|^2 \;
(\frac{1}{4}|\overline{f}|^2 + |\overline{\phi}|^2)
\nonumber \\[3mm] &+&
\frac{1}{2} g^2 (|\phi|^2-|\overline{\phi}|^2-
\frac{1}{2}(|f|^2-|\overline{f}|^2))^2\;,
\nonumber
\end{eqnarray}
%%%%%%%%%%%%%%%%
and possesses the exact D- and F- flat direction
$|\phi|=|\overline{\phi}|$ with $f=\overline{f}=0$.
Adding to $V_1$ the soft SUSY-breaking terms
%%%%%%%%%%%%%%%%%
\begin{eqnarray}
V_2 &=& m_f^2 |f|^2 + m_{\overline{f}}^2 |\overline{f}|^2+
m_{\phi}^2 |\phi|^2+m_{\overline{\phi}}^2 |\overline{\phi}|^2
\nonumber \\[3mm] &+& 
(\frac{1}{2} A_{\lambda} \lambda f f \phi +
\frac{1}{2} A_{\overline{\lambda}} \overline{\lambda} 
\overline{f} \overline{f} \overline{\phi}+
\frac{1}{2} M_1 \lambda_1 \lambda_1 + h.c\;),
\nonumber
\end{eqnarray}
%%%%%%%%%%%%%%%
the flat direction is lifted. However, $f=\overline{f}=0$
is still a minimum of $V=V_1+V_2$ for fixed $\phi$, $\overline{\phi}$
along the D-flat direction $|\phi|=|\overline{\phi}|$, provided
$|\phi|>>m_s$. Setting $f=\overline{f}=0$ and $|\phi|=|\overline{\phi}|>>m_s$
the potential $V$ reduces to 
%%%%%%%%%%%%%%%%
\begin{eqnarray}
V=(m_{\phi}^2 + m_{\overline{\phi}}^2 ) |\phi|^2\;.
\nonumber
\end{eqnarray}
%%%%%%%%%%%%%%%%%%%%%
A non-trivial minimum will occur at a scale $V_X\sim Q_0 >> m_s$ if
the quantity $m_{\phi}^2 (Q)+ m_{\overline{\phi}}^2(Q)$ flips
its sign at $Q=Q_0$, {\it i.e.}
%%%%%%%%%%%%%%%%%%%
%%%%%%%%%%%%%%%%
\begin{eqnarray}
m_{\phi}^2 (Q_0)+ m_{\overline{\phi}}^2(Q_0)=0\;.
\nonumber
\end{eqnarray}
%%%%%%%%%%%%%%%%%%%%%%%%
A more accurate determination of the position $V_X$ of the
minimum can be obtained numerically by solving the RGEs for
the soft SUSY-breaking terms and looking for a minimum at
$|\phi|=V_X$ of the ``effective potential" :
%%%%%%%%%%%%%
\begin{eqnarray}
(m_{\phi}^2 (|\phi|)+ m_{\overline{\phi}}^2(|\phi|))\;|\phi|^2 \;.
\nonumber
\end{eqnarray}
%%%%%%%%%%%%%%%%%
}}
, i.e.
%%%%%%%%%%%%%
\begin{equation}
m_{\cal{H}}^2(Q_0)+m_{\overline{\cal{H}}}^2(Q_0)=0 \;.
\label{condition}
\end{equation}
%%%%%%%%%%%%%
  Then, this reversal of sign signals the development
   of a symmetry breaking minimum
along the flat direction
  with v.e.v. $V_X\sim Q_0$ .

        The energy range in which $Q$ takes values is divided into three
regions:

      {\bf a) $M_X<Q<M_{s}$:}
The gauge couplings run according to the following RGEs
{\footnote {We ignore the Yukawa coupling contribution
to the 2-loop expression for the beta function of the gauge couplings.}}

%%%%%%%%%%%%%%%%
\begin{equation}
Q\frac{d\alpha_i}{dQ}=\frac{\alpha_i^2}{2\pi}(
b_i+\frac{1}{4\pi}\sum_{j=5,X}b_{ij}\alpha_j)
\label{eq:beta}
\end{equation}
%%%%%%%%%%%%%%%%%%%%%%

%%%%%%%%%%%%%%%%
\begin{eqnarray}
b_5&=&-2
\nonumber \\[3mm]
b_X&=&8
\end{eqnarray}
%%%%%%%%%%%%%%%%%%%%%%
and 
%\begin{equation}
\[b_{ij}= \left (
\begin{array}{clcr}
\frac{776}{5}   &       \frac{23}{5} \\ [3mm]
\frac{336}{5}   &       \frac{83}{10}
\end{array} \right) \;\;,(i,j=5,X).\]
%\end{equation}

{\bf b) $M_{10}<Q<M_X$ :}
In this energy region the RGEs receive contributions
from the spectrum of MSSM and from $D^c_{H'}({\overline{ 3}},1,1/3)$,
${\overline{ D}}^c_{\overline{ H'}}(3,1,-1/3)$, $Q_{H'}(3,2,1/6)$,
${\overline{ Q}}_{\overline{ H'}}({\overline{ 3}},2,-1/6)$ contained in $H'$ ,
${\overline{H}}'$ .
The 2-loop beta functions of the gauge couplings are given by
 $b_1=36/5$, $b_2=4$, $b_3=0$ and
%%%%%%%%%%%%%%%%%%%%%%%%%%%
\[b_{ij}= \left (
\begin{array}{clcr}
48   &    15     & 7 \\ [2mm]
40   &    46     & 2 \\ [2mm]
\frac{104}{5} & 6 & \frac{202}{25} \\[2mm]
\end{array} \right)\,\,\,\,, \,\,\, (i,j=3,2,1)\;. \]
%%%%%%%%%%%%%%%%%%%%%%%%%%%%%%%%
in a notation analogous to the one employed previously.
Here we evolve the Yukawa couplings as well as the soft SUSY-breaking masses
by making use of their one-loop beta functions.

{\bf c) $M_Z<Q<M_{10}$ :}

In this range there are contributions to the RGEs only from the MSSM spectrum.
For the evolution of all couplings and masses the 2-loop approximation was
made.

 In the following section we are going to combine the above RGEs in order to
achieve
 gauge coupling unification at $M_s\sim 5\times 10^{17} GeV$ and generation of
 the symmetry breaking scale $M_X$ in a way consistent with the low energy data.

{\bf 3. NUMERICAL ANALYSIS AND CONCLUSIONS.}

We evolve the 2-loop RGEs from $M_Z$
up to the scale
$M_{10}=3\times 10^{11}$ GeV , keeping
fixed the
 value for the strong coupling at $\alpha_s(M_Z)=0.120$
,  for $m_t=174$ GeV, $m_b=4.9$ GeV, $m_{\tau}=1.777$ GeV.
Above $M_{10}$ the thresholds of $H'$ , ${\overline {H}'}$ are
switched on.
The unification scale is defined by the equality of the gauge couplings
 $\alpha_3$ and $\alpha_2$ .
Above $M_X$ the
two gauge couplings $\alpha_5$ and $\alpha_X$ are evolved up to the
 string unification scale $M_s$ where
they
become equal. There,
we impose universal boundary conditions for the soft SUSY-breaking
masses
%%%%%%%%%%%%%%%
\begin{eqnarray}
m_{\cal H} &=& m_{\cal\overline{H}}=m_h=m_{\overline{h}}
=m_{{\cal F}_i}=m_{{f^c_i}}=m_{{\cal N}_i}\equiv M_0 \nonumber \\
M_1&=&M_5\equiv M_{1/2} \nonumber \\
A_{\lambda}&=&\overline{A_{\lambda}}=A_t=A_N^i\equiv A_0 \,\,\,
,\,\,\,\, i=1...3
\end{eqnarray}
%%%%%%%%%%%%%%%
and for simplicity we take the Yukawa couplings appearing in the superpotential
${\cal{W}}_1$
to be
%%%%%%%%%%%%%
\begin{equation}
\lambda={\overline {\lambda}}=Y_N^i\equiv\lambda_0 \;.
\end{equation}
%%%%%
Starting from these boundary conditions, we come down in energy and demand that
the relation
%%%%%%%%%%%%%
\begin{equation}
m_{\cal H}^2(Q_0)+m_{\cal\overline{H}}^2(Q_0)=0
\end{equation}
%%%%%%%%%%%%%%
be satisfied at a scale $Q_0\sim V_X \sim M_X<M_s$ .
The whole procedure is carried out
under the constraints
of electroweak radiative symmetry breaking at $M_Z$, perturbativity of all
couplings up to $M_s$ , as well as the
 experimental constraints on the values of $sin^2\theta$
and the sparticle masses \cite{DLT}.

%%%%%%%%%%%%%%%
\par
\vspace{.25in}
\begin{center}
{\hbox {\psfig{figure=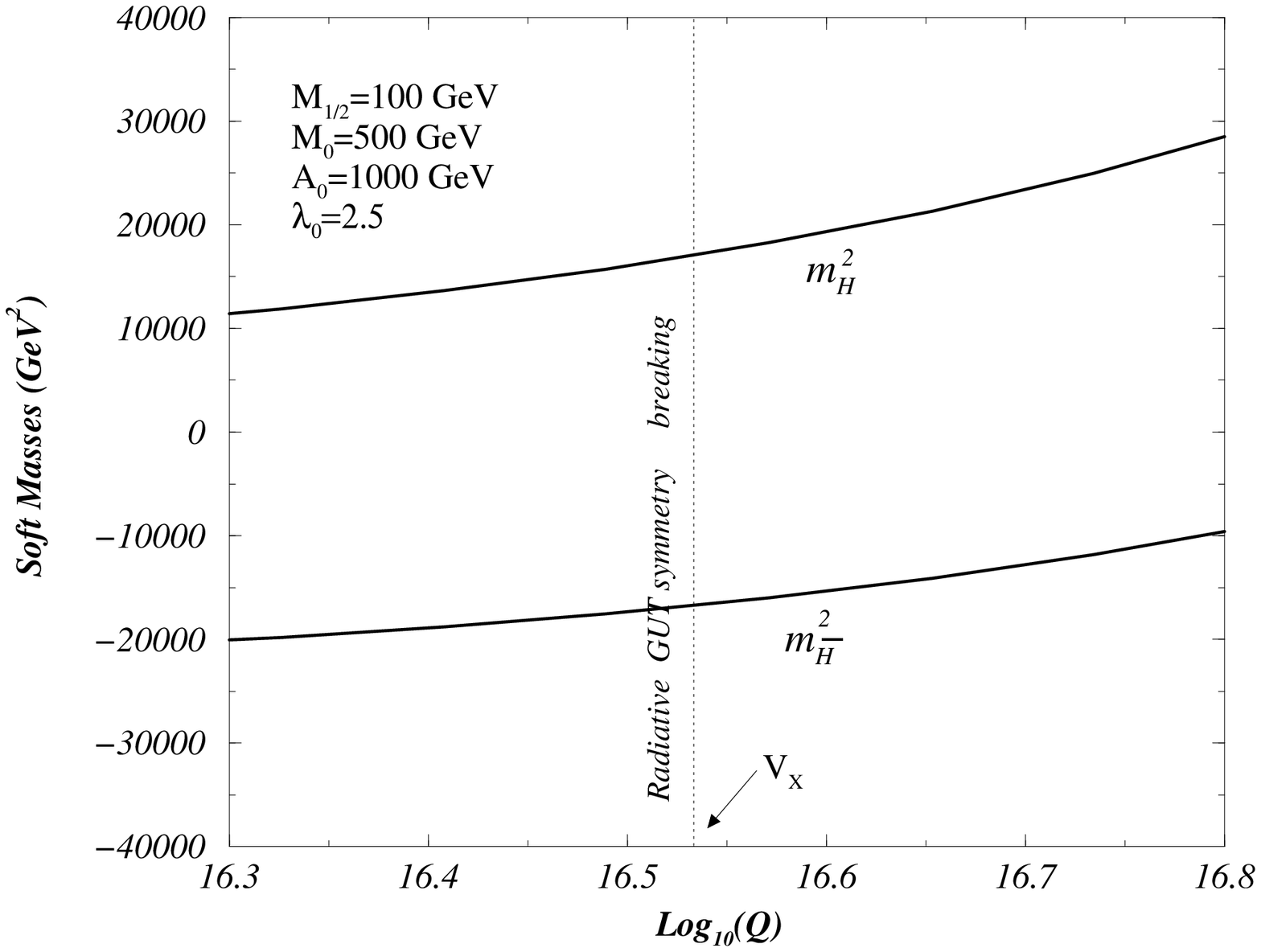,height=3in,width=3in}
\psfig{figure=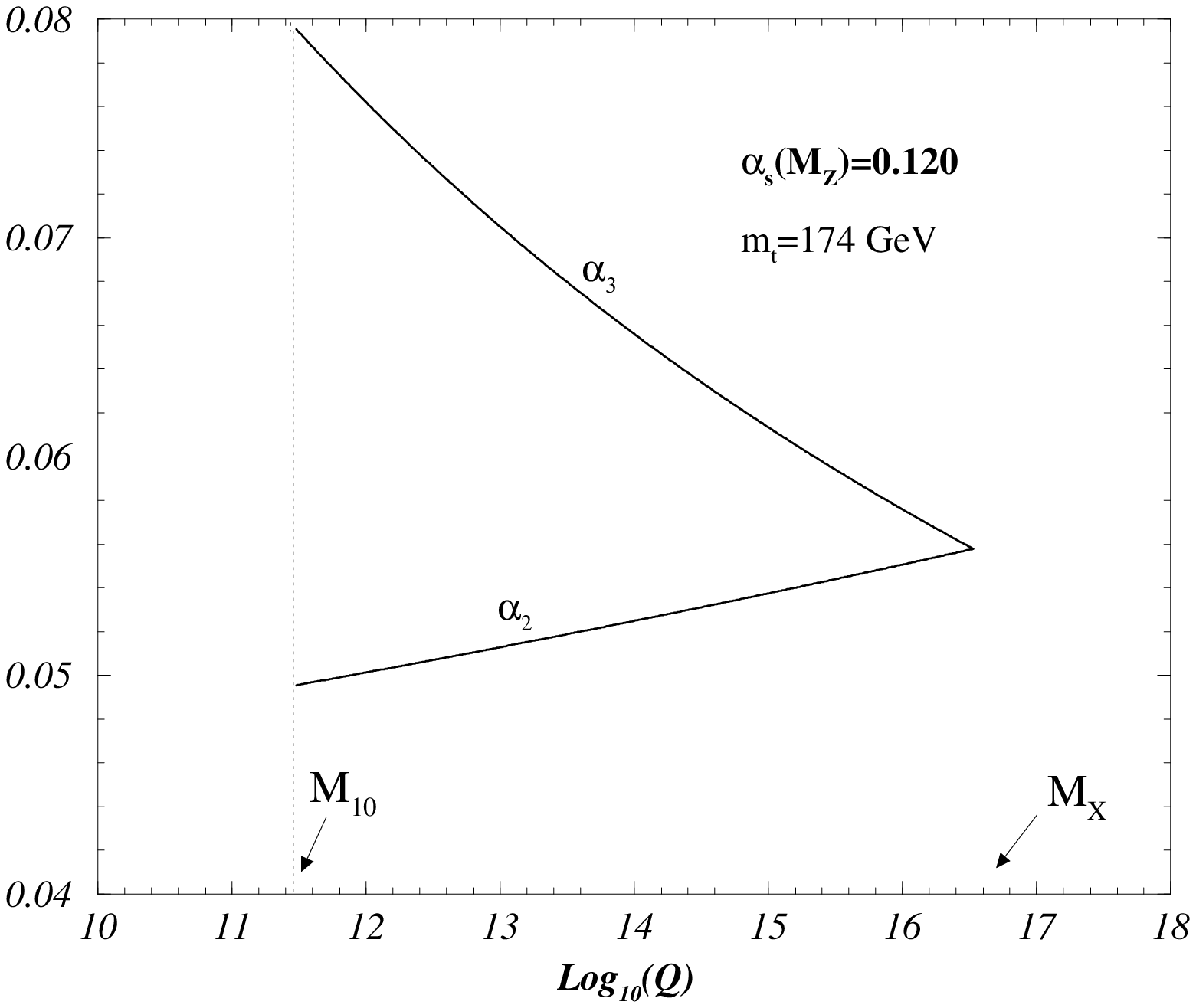,height=3in,width=3in}}}
\footnotesize{{\bf Figure 1:} Evolution of the
soft masses squared $m_{\cal H}^2$  and $m_{\cal\overline{H}}^2$
from $M_s\sim 5\times 10^{17}$ GeV
 to lower scales. The vertical line indicates the scale
$V_X$ where the radiative GUT symmetry
breaking of the flipped $SU(5)$ occurs. This scale practically
coincides
with the unification point of the gauge couplings
 $\alpha_2$ and $\alpha_3$.
}
\end{center}
\par
%\vspace{.25in}
%%%%%%%%%%%%%%%%%%%%%

The evolution of the soft masses squared $m_{\cal H}^2$ and
$m_{\cal\overline{H}}^2$ is depicted in Figure 1.
These masses although positive definite at $M_s$
develop a negative sign at a somewhat lower scale
 with the masses-squared of all the other
gauge non-singlets remaining positive , if we
adopt suitable values for the  parameters.
When the one
of the two  squared masses
(usually $m_{\cal\overline{H}}^2$)
becomes
negative, with its absolute value being greater than the value of the
 other, radiative GUT symmetry breaking occurs.
{}From the structure of the
renormalization group equations (\ref{soft1}),(\ref{soft2}),
it is easy to deduce  that
the way  to ensure the desired sequence of events is to keep
 gaugino masses $M_{1/2}$
at their
lowest possible value compatible with all the
 relevant phenomenological constraints and employ
values of $A_0$ and $M_0$ considerably larger than $M_{1/2}$.
Thus, for $M_0=500$ GeV, $A_0=1000$ GeV
, $M_{1/2}=100$ GeV and $\lambda_0=2.5$ the radiative
 GUT symmetry breaking picture outlined above is achieved.

 In Figure 1 we have taken  $M_{10}=3\times 10^{11}$ GeV.
If we increase $M_{10}$ by one order of magnitude,
$M_X$ decreases towards  a value  less than
$\simeq 10^{16}$ GeV, which is dangerous
for proton decay.
If we decrease $M_{10}$, the obtained value of
the  weak mixing angle
is quite large. The choice of the Yukawa
couplings $\lambda$=$\overline{\lambda}=Y_N^i=\lambda_0$
at $M_s$, plays a crucial
role in the determination of $M_X$ with the values
of $M_0$, $M_{1/2}$ and $A_0$ kept fixed.
In the region where $\lambda_0\lsim 2.0$, no radiative GUT symmetry breaking
occurs. On the other hand, if $\lambda_0\gsim 3.0$ the breaking takes place
at a scale which is too close to $M_s$ pushing among others the weak mixing
angle out of the  limits imposed by experiment.
Although the soft SUSY-breaking parameters  $A_0$ and $M_0$ must be
both large, in order to obtain the successful symmetry breaking picture just
outlined, they are still subject to the constraints of charge-color breaking
minima and radiative
electroweak symmetry breaking, respectively.
The case where $M_{1/2}$ is appreciable, and $M_0$, $A_0$ tend to
zero, results in $ M_X \sim M_s$ which
is unacceptable as already explained. In conclusion,
the generation of the superheavy scale is achieved in a relatively
narrow range of values of the parameters
under the assumption that $M_{1/2}<< M_0\,\,,\,\,A_0$ .

 The $SU(5)\times U(1)_X$ model studied in the present article
should be regarded
as a phenomenologically
viable example realizing
radiative GUT scale generation even in cases where
$F$-flatness is essentially exact.
The flatness of the potential is lifted by the small SUSY-breaking scalar mass 
squared terms which through radiative effects
flip their sign  along the almost flat
direction. The
relevant radiatively generated scale is
practically the energy at which this flipping
occurs.

{\bf Acknowledgments}

Two of us (A.D. and K.T.) acknowledge financial support from the research
program
$\Pi{\rm ENE}\Delta$-95
of the Greek Ministry
of Science and Technology. C.P. and K.T. acknowledge travelling support from the
TMR Network ``Beyond the Standard Model".

 \end{document}